\documentclass[article,12pt,onecolumn,showpacs,preprintnumbers,amsmath,assume,plain]{revtex4}
\usepackage{graphicx}
\DeclareGraphicsExtensions{.jpg}

\usepackage{dcolumn}
\usepackage{bm}
\usepackage{amsmath}
\usepackage{dcolumn}
\usepackage{bm}
\def\bk{ {\bf k} }

\def\bd{ {\bf d} }

\def\sr{ Sr$_2$RuO$_4$ }

\begin{document}

\title{Symmetry Field Breaking Effects in \sr}
\author{Pedro Contreras $^{1,\dag}$, Jos\'e Florez $^{2}$ and Rafael Almeida $^{3}$}
\affiliation{$^{1}$ Centro de F\'{\i}sica Fundamental, Universidad de Los Andes, M\'erida 5101, Venezuela\\
$^{2}$ Departamento de F\'{\i}sica, Universidad Distrital Francisco Jos\'e de Caldas, Bogot\'a , Colombia\\
$^{3}$ Departamento de Qu\'{\i}mica, Universidad de Los Andes,
M\'erida 5101, Venezuela\\
$^{\dag}$ Current address: Departamento de Ciencias B\'asicas,
Facultad de Ingenier\'{\i}a, Universidad Aut\'onoma de Caribe,
Barranquilla, Colombia.}

\date{\today}

\begin{abstract}
In this work, after reviewing the theory of the elastic properties
of \sr, an extension suitable to explain the sound speed experiments
of Lupien et. al. \protect\cite{lup2} and Clifford et. al.
\protect\cite{clif1} is carried out. It is found that the
discontinuity in the elastic constant C$_{66}$ gives unambiguous
experimental evidence that the \sr superconducting order parameter
$\Psi$ has two components and shows a broken time-reversal symmetry
state. A detailed study of the elastic behavior is performed by
means of a phenomenological theory employing the Ginzburg-Landau
formalism.


\end{abstract}


\maketitle

\section{Introduction}\label{sec:intro}

In a triplet superconductor the electrons in the Cooper pairs are
bound with spins parallel rather than antiparallel to one another,
i.e. they are bound in spin triplets \protect\cite{mae1,mac1,ric2}.
For this kind of superconductors, the spins are lying on the basal
plane, while the pair orbital momentum is directed along the
z-direction and their order parameter $\Psi$ is represented by a
three-dimensional vector $\bd (\bk)$. If $\Psi$ is of the type $k_x
\pm i\:k_y$, there is a Cooper pair residual orbital magnetism, which
gives place to an state of broken time reversal symmetry, edge
currents in the surface of the superconductor, and a tiny magnetic
field around non-magnetic impurities.

Based on the results of the Knight shift experiment performed
through the superconducting transition temperature T$_c$
\protect\cite{ish1,duf1}, it has been proposed that \sr is a triplet
superconductor. These experiments showed that Pauli spin susceptibility of the conduction electrons in the superconducting state
remains unchanged respect to its value in the normal state.
Moreover, it has been reported \protect\cite{luk1} that $\Psi$
breaks time reversal symmetry, which constitutes another key feature
of unconventionality.

The \sr elastic constants C$_{ij}$ have been measured as the
temperature T is lowered through T$_c$. The results show a
discontinuity in one of the elastic constants \protect\cite{lup2}.
This implies that
$\Psi$ has two different components with the time reversal symmetry
broken. Similar conclusions from a muon spin relaxation ($\mu$SR)
experiment were reported by Luke et. al. \protect\cite{luk1}.
Recently, experiments on the effects of uniaxial stress $\sigma_i$
, as a symmetry-breaking field, were performed by Clifford and
collaborators \protect\cite{clif1}, reporting that for \sr the
symmetry-breaking field can be controlled experimentally.
Additionally, experiments by Lupien et. al. \protect\cite{lup2}
showed the existence of small step in the transverse sound mode
T[100].

This body of results evidences the need of extending or developing
theoretical models to explain the changes occurring in the C$_{ij}$
at T$_c$, which, as far as we know, has not been carried out even in
quite recent works \protect\cite{clif1}. Thus, the aim of our work
is to extend an elasticity property phenomenological theory to show
that \sr is an unconventional superconductor with a two-component $\Psi$
\protect\cite{con1,con2}. Here, let us mention that a different theory of \sr
elastic properties was presented by Sigrist \protect\cite{sig2}.
However, unlike this paper, Sigrist work does not take into account
the splitting of T$_c$ due to $\sigma_i$, and directly calculates
the jumps at zero stress, where the derivative of T with respect to
$\sigma_i$ doesn't exist.

In this work, we first perform an analysis based on a $\Psi$ that
transforms as one of the two dimensional irreducible representations
of the \sr point group \protect\cite{con2,ric2}. Subsequently,
we construct the \sr superconducting phase diagram under an
external $\sigma_i$. This phase diagram is employed to develop
a complete theory of the
elastic behavior of \sr, based on a two component Ginzburg-Landau
($GL$) model. This theory allows to properly calculate the jumps in the
components of the elastic compliances S$_{ij}$. Finally, we propose
that there are significant advantages for using \sr as a material
for a detailed study of symmetry-breaking effects in
superconductivity described by a two-component $\Psi$.

\section{Ehrenfest relations for a uniaxial stress $\sigma_i$}

Provided that $\sigma_i$ does not split the phase transition
\cite{con2}, for applied $\sigma_i$, Ehrenfest relations
can be derived in analogous manner to the case of applied
hydrostatic pressure \cite{lan1,con1}, under the condition that T$_c$
is known as a function of $\sigma_i$.
In order to simplify the calculations, we make use of the Voigt
notation $i$ = $xx,yy,zz,yz,xz,xy$ \protect\cite{aul1}.

For a second order phase transition, the Gibbs
free energy $G$ derivatives respect to T, the entropy $S$ = $-(\partial
G/\partial T)_{\sigma}$, and respect to $\sigma_i$, the elastic strain e$_i$ = $-(\partial G/\partial \sigma_i)_T$ are
continuous functions of $\sigma_i$ and T. Therefore, at the
transition line, $\Delta e_i (T,\sigma_j)$ = 0 and $\Delta
S(T,\sigma_j)$ = 0. From this, for S and e$_i$, the boundary
conditions between the two phases are

\parbox{7cm}{\begin{eqnarray*}
\Delta \bigg[ \Big( \frac{\partial S}{\partial T} \Big)_{\sigma_j} \bigg]
dT +
\Delta \bigg[ \Big( \frac{\partial S}{\partial \sigma_j} \Big)_T \bigg]
d\sigma_j & = & 0 \\
\Delta \bigg[ \Big( \frac{\partial e_i}{\partial T} \Big)_{\sigma_j}
\bigg] dT +
\Delta \bigg[ \Big( \frac{\partial e_i}{\partial \sigma_j} \Big)_T \bigg]
d\sigma_i & = & 0
    \label{ehr_6}
\end{eqnarray*}}
\parbox{1cm}{\begin{eqnarray}\end{eqnarray}}

\noindent By using the definitions of the thermal expansion $\alpha_i$
= $(\partial e_i / \partial T)_{\sigma}$, the specific heat at
constant stress, \; $C_{\sigma}$ = $ T (\partial S /
\partial T)_{\sigma}$, and the elastic compliances S$_{ij}$ =
$(\partial e_i / \partial \sigma_j)_T$, together with the
Maxwell identity \; $(\partial S/\partial
\sigma_i)_T$ = $(\partial e_i/\partial T)_{\sigma_i}$, the previous
relations can be rewritten as,

\parbox{7cm}{\begin{eqnarray*}
\Delta \frac{C_\sigma}{T} + \frac{d \sigma_i}{d T} \; \Delta (\alpha_i)_{\sigma_i} & = & 0 \\
\Delta (\alpha_i)_{\sigma_j} +
\frac{d \sigma_j}{d T} \; \Delta ( \text{S}_{ij} )_{T} & = & 0.
    \label{ehr_66}
\end{eqnarray*}}
\parbox{1cm}{\begin{eqnarray}\end{eqnarray}}

\noindent From the first expression in eqn.~(\ref{ehr_66}), the
relation for $\alpha_i$ is found to be
\begin{equation}
    \Delta \: \alpha_i = - \Delta \: C_{\sigma} \; \frac{d
    \; \ln{T_c(\sigma_i)}}{d \sigma_i},
    \label{ehr_7}
\end{equation}
likewise, from the second expression of eqn.~(\ref{ehr_66}), the relation for S$_{ij}$ is obtained to be,
\begin{equation}
    \Delta \text{S}_{ij} = - \; \Delta \: \alpha_i \; \frac{d T_c(\sigma_j)}{d \sigma_j}.
    \label{ehr_8}
\end{equation}
\noindent It is important to distinguish that the print letter $S$ denotes
the entropy, while the symbol S$_{ij}$ means the elastic compliances.
In similar manner, the print letter $C$ stands for the specific heat
and the symbol C$_{ij}$ for the elastic stiffness. Let us also point out
that in deriving these expressions, we used the fact that for a
given thermodynamic quantity $Q$, its
discontinuity along the transition line points is obtained from
$\Delta Q = Q(T_c + 0^+) - Q(T_c - 0^+)$, where $0^+$ is a positive
infinitesimal quantity. Finally, by combining Eqs.~(\ref{ehr_7}) and ~(\ref{ehr_8}), the variation in S$_{ij}$ is found to be:
\begin{equation}
    \Delta \text{S}_{ij} = \frac{\Delta C_{\sigma}}{T_c}
    \: \frac{d T_c(\sigma_i)}{d \sigma_i}
    \: \frac{d T_c(\sigma_j)}{d \sigma_j}.
    \label{ehr_9}
\end{equation}

Before continuing, it is interesting to mention that besides of
our previous works \cite{con2,con1}, we are not aware of any other
works that have derived Ehrenfest relations for the case where
applied $\sigma_i$ produces a phase transition splitting.

\section{Ginzburg - Landau model} \label{dos}

In this section, a phenomenological model which takes into account
the \sr crystallographic point group D$_{4h}$ is derived and
employed. As we show, the analysis of $G$, using an order parameter
which belongs to any of the one dimensional representations of D$_{4h}$ is
not able to describe the splitting of T$_c$ under an external stress
field. In order to account properly for the splitting, superconductivity
in \sr must be described by a $\Psi$, transforming as one of the
D$_{4h}$ two dimensional irreducible representations, E$_{2g}$ or
E$_{2u}$, which at this level of theoretical description render
identical results \cite{con2,con1}.

\subsection{Superconducting free energy}

In order to derive a suitable $GL$ free energy
$G^{\Gamma}$, we first will suppose that the \sr superconductivity
is described by an order parameter $\psi^\Gamma$, which transforms
according to one of the eight one-dimensional representations of
D$_{4h}$: $\Gamma$ = A$_{1g}$, A$_{2g}$, B$_{1g}$, B$_{2g}$,
A$_{1u}$, A$_{2u}$, B$_{1u}$, or B$_{2u}$. Let us notice that
an analysis employing the D$_4$ point group renders similar results.
Here we will analyze the terms in $G^{\Gamma}$ linear in
$\sigma_i$ and quadratic in $\psi^\Gamma$:

\begin{eqnarray}
    G^\Gamma & = & G_0 + \alpha(T)|\psi^\Gamma|^2 + \frac{b}{2}\;|\psi^\Gamma|^4 +
    \nonumber \\
    & &
    [a\;(\sigma_{xx} + \sigma_{yy}) + c\;\sigma_{zz}] |\psi^\Gamma|^2.
\label{repres3}
\end{eqnarray}

The terms proportional to $\sigma_{xx}$, $\sigma_{yy}$ and
$\sigma_{zz}$ in eqn.~(\ref{repres3}) give rise to discontinuities
in the elastic constants, evidenced from sound speed measurements
\protect\cite{tes1}. On the other hand, discontinuities in the
elastic compliance S$_{66}$ and in the elastic constant C$_{66}$
\footnote{the Voigt notation for C$_{66}$ means C$_{xyxy}$ where $6$
= $xy$ \protect\cite{aul1}} arise from the linear coupling with
$\sigma_{xy}$. However, due to symmetry, the later linear coupling
does not exist for any $\Gamma$; therefore, S$_{66}$ and C$_{66}$ are
expected to be continuous at T$_c$ for any of the one-dimensional
irreducible representation that assumes a one-dimensional
$\psi^\Gamma$. Nevertheless, the results of Lupien et. al. experiments
 \protect\cite{lup2} showed a discontinuity in
C$_{66}$. Hence, based exclusively on sound speed
measurements, we conclude that none of the one-dimensional
irreducible representations can provide an appropriate description of
superconductivity in \sr . As far as we know, this conclusion has not been
previously established in the literature \protect\cite{clif1}. Let
us mention that for any one-dimensional $\Gamma$, a detailed
analysis of the calculation of the jumps in C$_{66}$ is
presented in ref.\protect\cite{con1}.

Due to the absence of discontinuity in S$_{66}$ for any of the
one-dimensional $\Gamma$, the superconductivity in \sr must be
described by an order parameter $\psi^E$ transforming as one of the
two-dimensional representations E$_{2g}$ or E$_{2u}$ \cite{con2}.
The $GL$ theory establishes that only the parameters
of one of the irreducible representations becomes non-zero at T$_c$.
Therefore, following the evidence provided in ref.~(\protect\cite{mae1,mae94}),
we choose the E$_{2u}$ spin-triplet state as the correct
representation for \sr, and the speed measurements are analyzed in terms
of the model $\psi^\text{E}$ $=$ $(\psi_x,\psi_y)$, with $\psi_x$ and
$\psi_y$ transforming as the components of a vector in the basal
plane. The expression for $G$ is determined by symmetry arguments
based on the analysis of the second and fourth order invariants
(real terms) of $G^{\Gamma}$. To maintain gauge symmetry, only real
and even products of $\Psi$ can occur in the expansion of
$G^{\Gamma}$; thus, we  find that all real invariants should be formed
by second and fourth order products of $\psi$`s \footnote{The
invariance under the gauge symmetry $U(1)$ means that the quantities
$\psi_i$ must transform according to the rule $\psi_x$ $\rightarrow$
$e^{i \Phi} \psi_x$ and $\psi_y$ $\rightarrow$ $e^{i \Phi} \psi_y$}.
To obtain its expression , we use the fact that $G$ is invariant
with respect to a transformation by the generators
c$_{4z}$ and c$_{2x}$ of D$_{4h}$.
Applying the generators to different second and fourth order
combination of products of $\psi$`s, we find only one second
order invariant $|\psi_x|^2 +|\psi_y|^2$ and three fourth order
invariants, namely
$|\psi_x|^2 |\psi_y|^2$, $|\psi_x|^4 +|\psi_y|^4$, and $\psi_x^2
\psi_y^{*2} + \psi_x^{*2} \psi_y^2$.

For the zero $\sigma_i$ case, the expansion of $G$ gives place to:
\begin{eqnarray}
    G  &=& G_0 + \alpha(T)(|\psi_x|^2
    +|\psi_y|^2) + \frac{b_1}{4}(|\psi_x|^2 +|\psi_y|^2)^2 +
    \nonumber \\
    & & b_2|\psi_x|^2 |\psi_y|^2 +
    \frac{b_3}{2}(\psi_x^2 \psi_y^{*2} + \psi_y^2 \psi_x^{*2}) ,
    \label{G_L_1}
\end{eqnarray}
where $\alpha$ = $\alpha^\prime (T - T_{c0})$ and the coefficients
b$_1$, b$_2$, and b$_3$ are material-dependent real constants
\protect\cite{sig1,min1}. These coefficients have to satisfy special
conditions in order to maintain the free energy stability. The
analysis of $G$ is accomplished by considering two component
$(\psi_x,\psi_y)$ with the form:
\begin{equation}
(\psi_x,\psi_y)= (\eta_x \hspace{0.2cm} e^{i \varphi/2}, i \; \eta_y \hspace{0.2cm} e^{-i \varphi/2}) ;
\label{GOPS}
\end{equation}
where $\eta_x$ and $\eta_y$ are both real and larger than zero.
After substitution of $\psi_x$ and $\psi_y$ in equation (7) \label{GOPS}, $G$ becomes:
\begin{eqnarray}
    G &=& G_0 + \alpha(T) (\eta_x^2
    +\eta_y^2) + \frac{b_1}{4}(\eta_x^2 +\eta_y^2)^2 +
    \nonumber \\
    & & (b_2 -b_3)\eta_x^2 \eta_y^2 + 2 b_3 \eta_x^2 \eta_y^2 \sin^2 \varphi.
    \label{G_L_2}
\end{eqnarray}

For fixed values of the coefficients b$_1$ and b$_2$, if b$_3$ $>$
0, $G$ will reach a minimal value if the last term vanishes, i.e. if
$\varphi$ = 0. Moreover, if $\eta_x$ and $\eta_y$ have the form
$\eta_x = \eta \hspace{0.2cm} \sin \chi$ and $\eta_y = \eta
\hspace{0.2cm} \cos \chi$, $G$ becomes
\begin{equation}
    G = G_0 + \alpha(T) \eta^2 + \frac{b_1}{4} \eta^4  -
\frac{\tilde{b}}{4}\eta^4 \sin^2 2\chi,
    \label{G_L_3}
\end{equation}
where $ \tilde{b} \equiv b_3 - b_2$. If $\tilde{b}$ $>$ 0, $G$
reaches its minimum value if $\sin^2 2\chi$ = 1, this condition
is satisfied if $\chi$ = $\pi/4$; and therefore $\eta_x$ = $\eta_y$.
On the other hand, if
$\tilde{b}$ $<$ 0, then $G$ becomes minimal if $\sin^2 2\chi$ = 0.
In this case, either $\eta_x$ = 0 or $\eta_y$ = 0. Since for a
superconducting state $(\psi_x,\psi_y)$ $\sim$ $(1,\pm i)$,
from the previous analysis, the
lowest $G$ state corresponds to $ b_3 - b_2 $ $>$ 0. This
thermodynamic state breaks time-reversal-symmetry; and hence, it is
believed to be the state describing superconductivity in \sr
\protect\cite{con2,mae1,mac1}. In addition, it is found that
for the phase transition to be of second order, it is required
that $b \equiv b_1 + b_2 - b_3$ $>$ 0.

At this point it is important to understand why the state
$(\psi_x,\psi_y)$ $\sim$ $(1,\pm i \epsilon)$ has been chosen for
the analysis of $\sigma_6$ and why it gives rise to the
discontinuity in S$_{66}$ \protect\cite{con1}. Minimization of
 eqn.~(7) with respect to $\varphi$ and $\chi$, and
employing eqn.~(8) renders a set of solutions for the two-component
order parameter  which depend on the relation between the
coefficients b$_1$, b$_2$, and b$_3$ and also on the value of the
phases $\varphi$ and $\chi$. Thus, for the $\text{E}$
representation, solutions of the form,
\begin{equation}
    \psi_1  = \eta \; (1,0) \; e^{i \; \varphi} ,
     \label{solut_OP_1}
\end{equation}
are obtained, which are very similar to those found for the D$_4$
one-dimensional irreducible representation.
Therefore, these solutions are not able to account for the
jump in C$_{66}$. However, solutions with both components
different than zero are also attained:
\begin{equation}
     \psi_2  = \frac{\sqrt{2}}{2} \: e^{i \; \pi/4} \: (1,1) \: \eta, \hspace{1.0cm}
     \psi_3  = \frac{\sqrt{3}}{2} \; (1,i) \eta.
     \label{solut_OP_2}
\end{equation}

\begin{figure}
\begin{center}
\includegraphics[width = 3.0 in, height= 2.5 in]{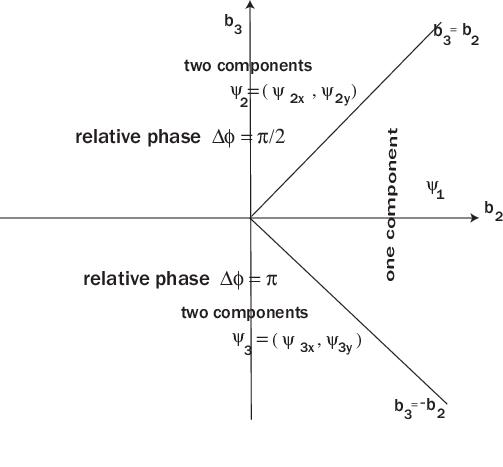}
\end{center}
\caption{\label{GL1} Superconducting state phase diagram for the
two dimensional representation $\text{E}$ of the tetragonal group
D$_4$ as function of the material parameters b$_2$ and b$_3$
showing the domains which correspond to the order parameters
$\psi_1$, $\psi_2$, and $\psi_3$. Each domain corresponds to a
different superconducting class.}
\end{figure}

Each of these solutions corresponds to different relations for the
$b_i$. This is illustrated by Fig.~(\ref{GL1}), which shows the
phase diagram, displaying the domains of
$\psi_1$, $\psi_2$ and $\psi_3$ as a function of $b_1$,
$b_2$ and $b_3$. Now, if the jump in C$_{66}$ corresponds to a $G$
minimum, the coupling term with $\sigma_6$ must be taken to be
different from zero. If the solution $\psi_2$ is considered, the
term containing $\sigma_6$ becomes zero; therefore it is not
acceptable. On the other hand, this requirement is satisfied by
$\Psi_3$, with the form $(1,i) \eta$. Hence, the $GL$
analysis renders $\Psi_3$ as the solution that breaks time reversal
symmetry.

\subsection{Coupling of the order parameter to an external stress}

The transition to an unconventional superconducting state shows
manifestations as the breakdown of symmetries, such as the crystal
point group or the time reversal symmetry  \protect\cite{sig1,min1}.
This loss of symmetry has measurable manifestations in observable
phenomena, as the splitting of T$_c$ under an elastic deformation.
The coupling between the crystal lattice and the superconducting
state is described Refs.~\protect\cite{sig1,min1}. As explained there,
close to T$_c$, a new term is added to $G$, which couples in second order
 $\Psi$ with e$_{ij}$ and in first order $\Psi$ with $\sigma_{ij}$.
These couplings give place to discontinuities in S$_{ijkl}$ at T$_c$.

\subsection{Analysis of the phase diagram}

An expression for $G$ accounting for a phenomenological coupling to
C$_{66}$ in the \sr basal plane is given by

{\begin{eqnarray}
    G  &=& G_0 + \alpha^\prime (T - T_{c0})(|\psi_x|^2
    +|\psi_y|^2) + b_2|\psi_x|^2 |\psi_y|^2 +
    \nonumber \\
    & & \frac{b_1}{4}(|\psi_x|^2 +|\psi_y|^2)^2 + \frac{b_3}{2}(\psi_x^2 \psi_y^{*2} + \psi_y^2 \psi_x^{*2}) -
    \nonumber \\
    & & \frac{1}{2} \: \text{S}_{ij} \: \sigma_i \: \sigma_j + \sigma_i \: \Lambda_i + \sigma_i \; d_{ij} \; E_j.
\label{ph_1}
\end{eqnarray}}
\
Here, $\Lambda_i$ are the temperature-dependent $\alpha_i$,
d$_{ij}$ are the coupling terms between $\Psi$ and S$_{ij}$
and $E_j$ are the invariant elastic compliance tensor components, defined below.
In order to determine these invariants describing the coupling
of the order parameter to the stress tensor, we
construct the tensor $E_j$ with Voigt components E$_1$ =
$|\psi_x|^2$, E$_2$ = $|\psi_y|^2$ and E$_6$ = $\psi_x^* \psi_y$ $+$
$\psi_x \psi_y^*$; where E$_6$  couples $\sigma_6$ and $\Psi$. The
tensor $d_{ij}$ couples $E_i$ with $\sigma_j$ and has the same
nonzero components as S$_{ij}$.
By applying symmetry considerations \protect\cite{con2}, it is shown
that the only non-vanishing independent components of d$_{ij}$ are
d$_{11}$, d$_{12}$ = d$_{21}$, d$_{31}$ = d$_{32}$, and d$_{66}$.
Contributions to $G$ that are quadratic in both, $\Psi$ and
$\sigma_6$ were neglected. Such terms would have given an additional
T dependence to the S$_{ij}$ \protect\cite{tes1}. However, given the
large number of independent constants occurring in the associated
sixth rank tensor, at this point, it is  not clear whether or not
the explicit inclusion of such terms would be productive.

\begin{figure}
\begin{center}
\includegraphics[width = 3.0 in, height= 2.0 in]{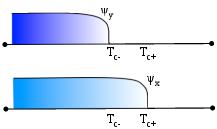}
\end{center}
\caption{\label{GL2} Temperature behavior of the two component order
parameter $(\psi_x,\psi_y)$ for the case of a nonzero uniaxial stress below
$T_c$. Notice that only the BCS component $\psi_x(T_{c+})$ becomes non zero
for temperatures between $T_{c+}$ and $T_{c-}$. The second unconventional component
$\psi_y(T_{c-})$ only appears  below $T_{c-}$.}
\end{figure}

Now, let us consider the case of uniaxial compression along the $a$
axis (only with $\sigma_1 < 0$). If in eqn.~(\ref{ph_1}), only
quadratic terms in $\Psi$ are kept, this equation can be written as

\begin{equation}
    G_{quad}  = \alpha^\prime [T - T_{c+}(\sigma_1)] |\psi_x|^2 +
    \alpha^\prime [T - T_{cy}(\sigma_1)] |\psi_y|^2,
\label{ph_2}
\end{equation}
here $T_{c+}(\sigma_1)$ and $T_{cy}(\sigma_1)$ are given by
\begin{equation}
    T_{c+}(\sigma_1)  = T_{c0} - \frac{\sigma_1 \: d_{11}}{\alpha^\prime}, \hspace{0.2cm}
    T_{cy}(\sigma_1) = T_{c0} - \sigma_1 \frac{d_{12}}{\alpha^\prime}.
    \label{Tc+}
\end{equation}

In what follows, we assume that $d_{11} - d_{12} >$ 0, such that
$T_{c+} > T_{cy}$. Notice that this does not imply any lost in
generality, assuming $d_{11} - d_{12} < 0$, would render an
identical model, simply by exchanging the $x$ and $y$ indices. Here,
T$_{c+}$ is the higher of the two critical
temperatures at which the initial
transition occurs. As should be expected, just below T$_{c+}$, only
$\psi_x$ is non zero. As T is further lowered, another phase
transition happens at T$_{c-}$, which is different than T$_{cy}$.
Below T$_{c-}$, the $\psi_y$ is also different from zero (see
fig.~(\ref{GL2})). Thus, in the presence of a non zero compressible
$\sigma_1$, $\Psi$ has the form $(\psi_x,\psi_y)$ $\approx$
$\psi(1,\pm i \, \epsilon)$, where $\epsilon$ is real and equal to
zero between T$_{c+}$ and T$_{c-}$ (phase 1), and increases from
$\epsilon$ = 0 to $\epsilon$ $\approx$ 1 as T becomes smaller than
T$_{c-}$ (phase 2), as illustrated in figs.~(\ref{GL1}) and
(\ref{GL2}).

The next step is finding T$_{c-}$. To achieve this goal, the equilibrium value of the non zero component of $\psi_x$, $\psi^2_x$ = $-2 \alpha_x/b_1$ is
replaced in eqn.~(\ref{ph_1}) and T$_{c-}$ follows from
\begin{equation}
T_{c+} - T_{c-} = - \Big[\frac{d_{11} - d_{12}}{2\,\alpha^{'}}\Big] \;
\Big[\frac{\tilde{b} + b}{\tilde{b}}\Big] \; \sigma_1.
\label{ph_3}
\end{equation}

\begin{figure}
\begin{center}
\includegraphics[width = 3.0 in, height= 2.0 in]{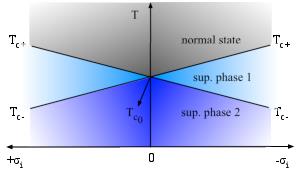}
\end{center}
\caption{\label{GL3} Phase diagram showing the upper and lower
superconducting transition temperatures, $T_{c+}$ and $T_{c-}$,
respectively, as functions of the compressible stress $-\sigma_i$
along the a axis.}
\end{figure}

To obtain eqn.~(\ref{ph_3}), it is assumed that $\sigma_1^2 \ll \sigma_1$
and only linear terms in $\sigma_1$ are kept. The phase diagram for this
system is shown in fig.~(\ref{GL3}).

\section{Calculation of the discontinuities} \label{tres}

As discussed before, an external uniaxial stress acting on the \sr
basal plane breaks the tetragonal symmetry of the crystal. As a
consequence of this, when a second order transition to the
superconducting state occurs, it splits into two transitions.
For the case of applied $\sigma_1$,
the analysis of the behavior of the sound speed at T$_c$
requires a systematic study of these second-order phase transitions.
Moreover, thermodynamic quantities, such as $dT_c/d
\sigma_i$, C$_{\sigma}$, and $\alpha_{\sigma}$, which are needed in
order to calculate the components S$_{ij}^{\sigma}$ are accompanied
by a discontinuity at each of the second order phase transitions.

\begin{figure}
\begin{center}
\includegraphics[width = 3.0 in, height= 2.0 in]{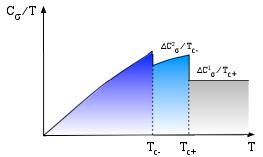}
\end{center}
\caption{\label{GL4} Schematic dependence of the specific heat on
the temperature, for the case of an uniaxial stress splitting the
Sr$_2$RuO$_4$ transition Temperature. Notice the two jumps in the
heat capacity near the transition temperatures T$_{c+}$ and
T$_{c-}$.}
\end{figure}

As depicted in fig.~(\ref{GL3}), for a given $\sigma_1 \neq 0$ as T
is lowered below T$_{c+}$, a first discontinuity for a thermodynamical quantity $Q$ is observed at the first line of transition
temperatures. This discontinuity along the transition
line, corresponding to the higher transition temperatures,
 $T$ = T$_{c+}$ $(\sigma_i)$ is given by $\Delta Q^+ = Q(T_{c+}
+ 0^+) - Q(T_{c+} - 0^+)$, where $0^+$ is a positive infinitesimal
number. If T is further dropped below T$_{c-}$, a second
discontinuity arises, and the lower line of transition
temperatures appears. The
discontinuity along this line, at T = $T_{c-}$ $(\sigma_i)$, is
defined by $\Delta Q^- = Q(T_{c-} + 0^+) - Q(T_{c-} - 0^+)$
\protect\cite{walkchr}. The sum of these two discontinuities
\begin{equation}
\Delta Q (T_{c0},\sigma=0) =\Delta Q^+ + \Delta Q^-,
\label{nuevecita}
\end{equation}
gives the correct expressions for the discontinuities at T$_{c0}$,
for the case with $\sigma_i$= 0, where the Ehrenfest relations do not
hold directly \protect\cite{con2}. As an example of these
discontinuities, the two jumps in C$_{\sigma}$
under an external $\sigma_i$ are sketched in fig.~(\ref{GL4}).

\subsection{Jumps due to a uniaxial stress $\sigma_1$}

The free energy, eqn.~(\ref{ph_1}), for the cases where both $\sigma_1$ and
$\sigma_6$ are nonzero is:

\begin{eqnarray}
    G  &=& G_0 + \alpha_x |\psi_x|^2 + \alpha_y |\psi_y|^2
    + \sigma_6 d_{66} (\psi_x \psi^{\ast}_y + \psi^{\ast}_x \psi_y) +
    \nonumber \\
    & & \frac{b_1}{4}(|\psi_x|^2 +|\psi_y|^2)^2 + b_2|\psi_x|^2 |\psi_y|^2 +
    \nonumber \\
    & & \frac{b_3}{2}(\psi_x^2 \psi_y^{*2} + \psi_y^2 \psi_x^{*2}).
    \label{st_1}
\end{eqnarray}

Here $\alpha_x$ = $\alpha^\prime (T - T_{c0}) + \sigma_1 d_{11}$ and
$\alpha_y$ = $\alpha^\prime (T - T_{c0}) + \sigma_1 d_{12}$. If only
$\sigma_1$ is applied, this equation becomes:
\begin{eqnarray}
    \Delta G  &=& \alpha_x |\psi_x|^2 + \alpha_y |\psi_y|^2 +
    \frac{b_1}{4}(|\psi_x|^2 +|\psi_y|^2)^2 +
    \nonumber \\
    & & b_2|\psi_x|^2 |\psi_y|^2 + \frac{b_3}{2}(\psi_x^2 \psi_y^{*2} + \psi_y^2 \psi_x^{*2}),
\label{st_6}
\end{eqnarray}
where $\Delta G = G - G_0(T)$. The nature of the superconducting
state that follows from eqn.~(\ref{st_6}), depends on the values of
the coefficients b$_1$, b$_2$, and b$_3$. The analysis from
eqn.~(\ref{st_6}) of the superconducting part of $G$ is
performed by using, as was done previously, an expression for $\Psi$
given by eqn.~(8).

At T$_{c+}$ and in the presence of $\sigma_1$, the second order
terms in eqn.~(\ref{st_6}) dominate and $\Psi$ has a single
component $\psi_x$; whereas at T$_{c-}$ a second component $\psi_y$
appears. Thus, at very low T, the fourth order terms dominate
the eqn.~(\ref{st_6}) behavior. Each of these two-component
domains has the form of $\psi_2$ given by eqn.~(\ref{solut_OP_2}).
In this case, $G$ can be written in terms of $\eta_x$ and $\eta_y$
as

\begin{eqnarray}
    \Delta G &=& \alpha_x \eta_x^2 + \alpha_y \eta_y^2 +
    \frac{b_1}{4}(\eta_x^2 +\eta_y^2)^2 +
    \nonumber \\
    & & (b_2 - b_3)\eta_x^2 \eta_y^2 + 2 b_3 \eta_x^2 \eta_y^2 \sin^2 \varphi.
    \label{st_8}
\end{eqnarray}

The analysis of eqn.~(\ref{st_8}) depends on the relation between
the coefficients b$_1$, b$_2$, and b$_3$. Assuming that b$_3$ $>$ 0,
and $\eta_x$ and $\eta_y$ are both different from zero, and
following the procedure described after eqn. (9) one arrives to

\begin{equation}
(\psi_x,\psi_y) \approx (1,\pm i \epsilon),
\end{equation}
where $\epsilon$ is real and grows from $\epsilon$ = 0 to $\epsilon$
$\approx$ 1 as T is reduced below T$_{c-}$, while eqn.~(\ref{st_8})
becomes

\begin{equation}
    \Delta G = \alpha_x \eta_x^2 + \alpha_y \eta_y^2 +
    \frac{b_1}{4}(\eta_x^2 + \eta_y^2)^2  - (b_3 - b_2) \eta_x^2 \eta_y^2.
    \label{st_9}
\end{equation}

To calculate the jumps at T$_{c+}$, we use $\alpha_x$ =
$\alpha^\prime (T - T_{c+})$ and $\alpha_y$ = $\alpha^\prime (T -
T_{cy})$, and assume that T$_{c+}$ $>$ T$_{cy}$. For the interval
$T_{c+} > T > T_{c-}$, the equilibrium value for $\Psi$ satisfies
$\alpha_x > 0$ and $\alpha_y = 0$, i.e. $\eta_x$ $>$ 0 and $\eta_y$
= 0, with $\eta^2_x = -2 \alpha_x/b_1$, obtaining that T$_{c+}$ and its derivative with respect to $\sigma_1$ are respectively,

\parbox{7cm}{\begin{eqnarray*}
T_{c+}(\sigma_1)  & = & T_{c0} - \frac{\sigma_1}{\alpha^\prime} \; d_{11}, \\
\frac{d \; T_{c+}}{d \; \sigma_1} & = &
-\frac{d_{11}}{\alpha^\prime}.
\label{st_10}
\end{eqnarray*}}
\parbox{1cm}{\begin{eqnarray}\end{eqnarray}}

\noindent The specific heat discontinuity at T$_{c+}$, relative to its
normal state value, is calculated by using:
\begin{equation}
\Delta C_{\sigma_1} = - T \; \frac{\partial^2 \Delta G}{\partial^2 T} \; \vert_{T = T_{c+}},
\end{equation}
and renders the result
\begin{equation}
    \Delta C_{\sigma_1}^+ = - \frac{2 \; T_{c+} \alpha^{\prime2}}{b_1}.
\label{st_11}
\end{equation}

A schematic depiction of the C$_{\sigma}$ discontinuities below
this transition temperature is exhibited in fig.~(\ref{GL4}).
At T$_{c+}$, the discontinuity in $\alpha_{\sigma}$ is
calculated by applying the
Ehrenfest relation of eqn.~(\ref{ehr_7}), yielding:
\begin{equation}
\Delta \alpha_{1}^{+}  =  - \frac{2 \alpha^\prime \; d_{11}}{b_1}.
\label{st_12}
\end{equation}

The discontinuities in S$_{ij}$ are obtained by using
eqns.~(\ref{ehr_8}) and ~(\ref{ehr_9}), rendering the result,
\begin{equation}
\Delta S_{i^\prime j^\prime}^{+}  = - \frac{2 \; d_{i^\prime 1} \; d_{j^\prime 1}}{b_1}.
\label{st_13}
\end{equation}

In the previous expression a prime on an index (as in $i^\prime$ or $j^\prime$) indicates a Voigt index taking only the values 1,2, or 3. Thus, from eqn. (27) the change in S$_{11}$ at T$_{c+}$ can be calculated to be $\Delta S_{11}^{\sigma_1} = - \frac{2 \; d_{11}^{2}}{b_1}$.

To find the discontinuities at T$_{c-}$, the term $(\eta_x^2 + \eta_y^2)^2$ in eqn.~(\ref{st_9}) is expanded, after which $G$
takes the form,

\begin{equation}
    \Delta G = \alpha_x \eta^2_x + \frac{b_1}{4} \eta^4_x +
    \Big[\alpha_y + (\frac{b_1}{2}+ b_2 - b_3) \eta_x^2\Big] \eta^2_y + \frac{b_1}{4} \eta^4_y.
    \label{st_14}
\end{equation}

In this expression, the second order term in $\eta_y$ is
renormalized by the square of $\eta_x$. The second transition temperature is determined from the zero of the total prefactor of $\eta^2_y$, obtaining that T$_{c-}$ and its derivative with respect to $\sigma_1$ are:

\parbox{7cm}{\begin{eqnarray*}
T_{c-}(\sigma_1)  & = & T_c -
\frac{\sigma_1}{2\alpha^\prime}\Big[d_{11} + d_{12} - \frac{b}{\tilde{b}}(d_{12} - d_{11})\Big], \\
\frac{d \; T_{c-}}{d \; \sigma_1} & = & -\frac{1}{2 \, \alpha^\prime}
\Big[d_{11} + d_{12} - \frac{b}{\tilde{b}}(d_{12} - d_{11})\Big].
\label{st_15}
\end{eqnarray*}}
\parbox{1cm}{\begin{eqnarray}\end{eqnarray}}

\noindent Below T$_{c-}$ the superconducting free energy,
eqn.~(\ref{st_14}) has to be minimized respect to both
components of $\Psi$. After doing so, $\eta_x$ and $\eta_y$
for this temperature range are found to be

\parbox{7cm}{\begin{eqnarray*}
\eta^2_x & = & - \frac{1}{2 \; b \; \tilde{b}}\Big[ (b - \tilde{b})\alpha_y +
(b + \tilde{b})\alpha_x\Big], \\ \eta^2_y & = & - \frac{1}{2 \; b \;
\tilde{b}}\Big[ (b - \tilde{b})\alpha_x + (b -\tilde{b})\alpha_y\Big].
\label{st_16}
\end{eqnarray*}}
\parbox{1cm}{\begin{eqnarray}\end{eqnarray}}

\noindent This analysis shows that the second superconducting phase
is different in symmetry, and that time reversal symmetry is
broken. The change in $C_{\sigma_1}$ at T$_{c-}$, with respect to
its value in the normal phase, $\Delta C_{\sigma_1}^{-,N}$, is found
to be, $\Delta C_{\sigma_1}^{-,N}$ = $- 2 \; T_{c-}\alpha^{\prime}
\: 1/b$. The specific heat variation at T$_{c-}$ is,
\begin{equation}
\Delta C_{\sigma_1}^{-} = \Delta C_{\sigma_1}^{-,N} - \Delta C_{\sigma_1}^{+},
\end{equation}
which results in
\begin{equation}
    \Delta C_{\sigma_1}^{-}  =  - 2 \; T_{c-} \; \alpha^{\prime 2}
        \frac{\tilde{b}}{b \; b_1}.
\label{st_17}
\end{equation}

The size of these jumps is complicated to infer, because it depends
on the material parameters b$_1$, b$_2$ and b$_3$, and on the
coupling constants d$_{11}$ and d$_{12}$.

With the help of the Ehrenfest relation, eqn.~(\ref{ehr_7}), the
discontinuity in $\alpha_i$ at T$_{c-}$ is obtained to be
\begin{equation}
    \Delta \alpha_{i^\prime}^{-} = - \alpha^\prime \;
\frac{\tilde{b}}{b \; b_1} \; \Big(d_{i^\prime +} - \frac{b}{\tilde{b}} \; d_{i^\prime -}\Big),
\label{st_18}
\end{equation}
and after employing eqns.~(\ref{ehr_8}) and (\ref{ehr_9}), the
discontinuity in $\text{S}_{i^\prime,j^\prime}$ at T$_{c-}$ is shown to be
\begin{equation}
    \Delta \text{S}_{i^\prime j^\prime }^{-} = - \frac{\tilde{b}}{2 \; b \; b_1} \;
        (d_{i^\prime +} - \frac{b}{\tilde{b}} d_{i^\prime -}) \;
        (d_{j^\prime +} - \frac{b}{\tilde{b}} d_{j^\prime -}).
\label{st_19}
\end{equation}

Here d$_{i^\prime \pm}$ = d$_{i^\prime 1} \pm d_{i^\prime 2}$. The
discontinuities occurring at T$_{c0}$, in the absence of uniaxial stress,
can be obtained by adding the discontinuities occurring at T$_{c+}$ and
T$_{c-}$, yielding:

\begin{eqnarray}
    \Delta C_{\sigma_1}^{0} & = & - \frac{2 T_{c0} \alpha^{\prime 2}}{b},
        \hspace{0.4cm} \Delta \alpha_{i^\prime}^{0} =
        - \frac{ \alpha^\prime d_{i^\prime +}}{b}, \nonumber \\
    \Delta \text{S}_{i^\prime j^\prime }^0 & = & - \frac{1}{2}
        \bigg(\frac{d_{i^\prime +} \; d_{j^\prime +}}{b} +
\frac{d_{i^\prime -} \;
        d_{j^\prime -}}{\tilde{b}}\bigg).
    \label{st_20}
\end{eqnarray}

Before continuing, it is important to emphasize that at at zero stress,
the derivative of T$_c$ with respect to $\sigma_i$ is not defined;
therefore, there is no reason to expect any of the Ehrenfest
relations to hold.

\subsection{Jumps due to a shear stress $\sigma_6$}

When a shear stress $\sigma_6$ is applied to the basal plane of \sr,
the crystal tetragonal symmetry is broken, and a second order
transition to a superconducting state occurs. Accordingly, for this case
the analysis of the sound speed behavior at T$_c$ also requires a
systematic study of the two successive
second order phase transitions. Hence, the C$_{66}$ discontinuity
observed by Lupien et. al. \protect\cite{lup2} at T$_c$, can be
explained in this context.

If there is a double transition, the derivative of
T$_c$ with respect to $\sigma_6$ i.e. $d T_c/d \sigma_6$ is
different for each of the two transition lines. At each of these
transitions, C$_{\sigma_6}$, $\alpha_{\sigma_6}$, and
S$_{ij}^{\sigma_6}$ show discontinuities. As discussed before,
the sum of them gives the correct expressions for the
discontinuities at zero shear stress, where the Ehrenfest relations
do not hold.

The T$_c$-$\sigma_6$ phase diagram will be similar to that obtained
for $\sigma_1$; therefore, the diagram in fig.(\ref{GL3}) also
qualitatively holds here. In the case of an
applied $\sigma_6$, $\Delta G$ is given by
\begin{eqnarray}
   \Delta G & = &\alpha ( |\psi_x|^2 + |\psi_y|^2 ) + \sigma_6 d_{66} (\psi_x \psi^{\ast}_y + \psi^{\ast}_x \psi_y) +
   \nonumber\\
   & & \frac{b_1}{4}(|\psi_x|^2 +|\psi_y|^2)^2 + b_2|\psi_x|^2 |\psi_y|^2 +
   \nonumber \\
   & & \frac{b_3}{2}(\psi_x^2 \psi_y^{*2} + \psi_y^2 \psi_x^{*2}).
   \label{sts_1}
\end{eqnarray}

Here $\alpha$ = $\alpha^\prime (T - T_{c0})$, and the minimization
of $\Delta G$ is performed as in  the $\sigma_1$ case,
i.e. by substituting the general expression for $\Psi$
given in eqn.~(8). After doing so, $\Delta G$ becomes
\begin{eqnarray}
    \Delta G & = & \alpha (\eta_x^2 + \eta_y^2) + 2 \eta_x \eta_y \: \sigma_6 \sin \varphi
    \: d_{16} + \frac{b_1}{4}(\eta_x^2 +\eta_y^2)^2 + \nonumber \\
    & & (b_2 - b_3)\eta_x^2 \eta_y^2 + 2 b_3 \eta_x^2 \eta_y^2 \sin^2 \varphi.
    \label{st_ss3}
\end{eqnarray}

In the presence of $\sigma_6$, the second order term determines the
phase below T$_{c+}$, which is characterized by $\psi_x$ and by
$\psi_y$ $=$ 0. As the temperature is lowered below T$_{c-}$,
depending of the value of b$_3$ a second component $\psi_y$
may appear. If at T$_{c-}$ a second component occurs, the fourth
order terms in eqn.~(\ref{st_ss3}) will be the dominant one.
Thus for very low T's, or for $\sigma_6$ $\rightarrow$ 0,
a time-reversal symmetry-breaking superconducting state
may emerge. The analysis of eqn.~(\ref{st_ss3}) depends
on the relation between the coefficients
b$_2$ and b$_3$. It also depends on the values of the quantities
$\eta_x$ and $\eta_y$, and of the phase $\varphi$. If b$_3$ $<$ 0,
and $\eta_x$ and $\eta_y$ are both nonzero, the state with minimum
energy has a phase $\varphi$ = $\pi/2$. The transition temperature
is obtained from eqn.~(\ref{st_ss3}), by performing the canonical transformations:
$\eta_x$ = $\frac{1}{\sqrt{2}}(\eta_{\mu} + \eta_{\xi})$ and
$\eta_y$ = $\frac{1}{\sqrt{2}}(\eta_{\mu} - \eta_{\xi})$.
After their substitution, eqn.~(\ref{st_ss3}) becomes
\begin{equation}
    \Delta G = \alpha_{+} \eta_{\xi}^2 + \alpha_{-} \eta_{\mu}^2  +
    \frac{1}{4}(\eta_{\xi}^2 + \eta_{\mu}^2)^2 + (b_2 + b_3)(\eta_{\xi}^2 - \eta_{\mu}^2)^2.
    \label{st_ss5}
\end{equation}

If, as was done before, $\eta_{\xi} = \eta \hspace{0.1cm} \sin \chi$
and $\eta_{\mu} = \eta \hspace{0.1cm} \cos \chi$, eqn.~(\ref{st_ss5})
takes the form
\begin{equation}
    \Delta G = \alpha_{+} \eta^2 \sin^2\chi + \alpha_{-} \eta^2 \cos^2\chi  +
    \frac{\eta^4}{4}\Big[ b_1 + (b_2 + b_3) \cos^2 2\chi \Big].
    \label{st_ss6}
\end{equation}

$\Delta G$ is minimized if $\cos 2\chi $ = 1, this is, if $\chi$
$=$ 0. Also, in order for the phase transition to be of second
order, $b^\prime$, defined as $b^\prime \equiv b_1 + b_2 + b_3$,
must be larger than zero. Therefore, if $\sigma_6$ is non zero, the
state with the lowest free energy corresponds to $b_3$ $<$ 0,
phase $\varphi$ equal to $\pi/2$, and $\Psi$ of the form:
\begin{equation}
(\psi_x,\psi_y) \approx \eta \: (e^{\frac{i \varphi}{2}}, \: e^{-\frac{i \varphi}{2}}).
\end{equation}

In phase 1 of fig.~(\ref{GL3}), $\varphi$ = 0, and as T is lowered
below T$_{c-}$, phase 2, $\varphi$ grows from 0 to approximately
$\pi/2$. Again, following an analysis similar to that carried out for
$\sigma_1$, the two transition temperatures T$_{c+}$ and T$_{c-}$
are obtained to be:

\parbox{7cm}{\begin{eqnarray*}
T_{c+}(\sigma_6) & = & T_{c0} - \frac{\sigma_6}{\alpha^\prime} d_{66}, \\
T_{c-}(\sigma_6) & = & T_{c0} + \frac{b}{2 \: b_3 \: \alpha^\prime} \sigma_6 \; d_{66}  .
\label{st_s7}
\end{eqnarray*}}
\parbox{1cm}{\begin{eqnarray}\end{eqnarray}}

\noindent The derivative of T$_{c+}$ with respect
to $\sigma_6$, and the discontinuity in $C^+_{\sigma_{6}}$
at T$_{c+}$  are respectively found to be:

\parbox{7cm}{\begin{eqnarray*}
\frac{d \; T_{c+}}{d \; \sigma_6} & = & - \frac{d_{66}}{\alpha^\prime}, \\
\Delta C^+_{\sigma_{6}} & = & -\frac{2 \: T_{c+} \: \alpha^{\prime 2}}{b^\prime}.
\label{st_s81}
\end{eqnarray*}}
\parbox{1cm}{\begin{eqnarray}\end{eqnarray}}

\noindent After applying the Ehrenfest relations,
eqns.~(\ref{ehr_8}) and (\ref{ehr_9}), the results for $\Delta \alpha_{\sigma_6}$ and $\Delta \text{S}_{66}$ at T$_{c+}$ are:

\parbox{7cm}{\begin{eqnarray*}
\Delta \alpha_{\sigma_{6}}^+ & = & -\frac{2 \: \alpha^\prime \: d_{66}}{b^\prime}, \\
\Delta \text{S}_{66}^+ & = & -\frac{2 \: d_{66}^2}{b^\prime}.
\label{st_s82}
\end{eqnarray*}}
\parbox{1cm}{\begin{eqnarray}\end{eqnarray}}

\noindent For T$_{c-}$, the derivative of this transition temperature with respect to $\sigma_6$, and the discontinuities in the specific heat,
thermal expansion and elastic stiffness respectively are:

\parbox{7cm}{\begin{eqnarray*}
\frac{d \; T_{c-}}{d \; \sigma_6} & = &  \frac{b \: d_{66}}{2 \: b_3 \: \alpha^\prime}, \\
\Delta C_{\sigma_{6}}^- & = & -\frac{4 \: T_{c-} \: \alpha^{\prime 2} \: b_3}{b \: b^\prime},
\label{st_s91}
\end{eqnarray*}}
\parbox{1cm}{\begin{eqnarray}\end{eqnarray}}

\parbox{7cm}{\begin{eqnarray*}
\Delta \alpha_{\sigma_{6}}^- & = & \frac{2 \: \alpha^\prime \: d_{66}}{b^\prime}, \\
\Delta \text{S}_{66}^- & = & -\frac{d_{66}^2 \: b}{b^\prime b_3}.
\label{st_s92}
\end{eqnarray*}}
\parbox{1cm}{\begin{eqnarray}\end{eqnarray}}

\noindent

Since for the case of $\sigma_6$, the derivative of T$_c$ with
respect to $\sigma_6$ is not defined at zero stress point,
the Ehrenfest relations do not hold at T$_{c0}$.
Thus, the discontinuities occurring at T$_{c0}$,
in the absence of $\sigma_6$, are calculated
by adding the expressions obtained for the discontinuities at
T$_{c+}$ and T$_{c-}$,

\parbox{7cm}{\begin{eqnarray*}
\Delta C_{\sigma_{6}}^0  & = &  -\frac{2 \: T_{c0} \: \alpha^{\prime 2}}{b}, \\
\Delta \text{S}_{66}^0 & = & -\frac{d_{66}^2}{b_3}, \\
\Delta \alpha_{\sigma_{6}}^0  & = &  0.
\label{st_s10}
\end{eqnarray*}}
\parbox{1cm}{\begin{eqnarray}\end{eqnarray}}

\noindent Notice that in this case, there is no discontinuity for
$\alpha_{\sigma_{6}}^0$.

Since the phase diagram was determined as a function of $\sigma_6$,
rather than as a function of the strain, (see fig.~(\ref{GL3})),
in this work, as in refs.~\protect\cite{con2,con1}, we make use of
the 6 $\times$ 6 elastic compliance matrix $\text{S}$, whose matrix
elements are S$_{ij}$. However, the sound speed measurements are best
interpreted in terms of the elastic stiffness matrix $\text{C}$,
with matrix elements C$_{ij}$, which is the inverse of $\text{S}$
\protect\cite{nei1}. Therefore, it is important to be able to
obtain the discontinuities in the elastic stiffness matrix in terms
of the elastic compliance matrix. Thus, close to the transition
line, $\text{C}(T_c + 0^+)$ = $\text{C}(T_c - 0^+) + \Delta \;
\text{C}$ and $\text{S}(T_c + 0^+)$ = $\text{S}(T_c - 0^+) + \Delta
\text{S}$, where $0^+$ is positive and infinitesimal. By making use
of the fact that $\text{C}(T_c + 0^+) \; \text{S}(T_c + 0^+)$ =
$\hat{1}$, where $\hat{1}$ is the unit matrix, it is shown that, to
first order, the discontinuities satisfy, $\Delta \text{C}$
$\approx$ $- \text{C} \: \Delta \text{S} \: \text{C}$. In this
manner, it is found that, for instance at T$_{c+}$, $\Delta \text{C}_{11}^{+}$ $\approx$ $\frac{2
\; (\text{C}_{j1} \; d_{j1})^2}{b_1}$. From this expression it is
clear that $\Delta \text{C}_{11}^{+}$ must be greater than zero.
In general, at T$_{c+}$, T$_{c-}$, and T$_{c0}$, the expressions
that define the jumps for the discontinuities in elastic
stiffness and compliances, due to an external stress,
have either a positive or a negative value. In this way,
$\Delta\text{S}_{11}$, $\Delta \text{S}_{22}$, $\Delta \text{S}_{33}$,
and $\Delta \text{S}_{66}$ are all negative; while, the stiffness
components $\Delta \text{C}_{11}$, $\Delta \text{C}_{22}$, $\Delta
\text{C}_{33}$, and $\Delta \text{C}_{66}$ are all positive.

\section{Final remarks} \label{cuatro}

Since for \sr, the symmetry-breaking field, due to $\sigma_i$, is under
experimental control, states of zero symmetry-breaking stress and of
$\sigma_i$ single direction can be achieved
\protect\cite{lup1,lup2,clif1}. Hence, it has
significant advantages the use of \sr as a material
in detailed studies of superconductivity symmetry-breaking
effects, described by a two-component order parameter.
Nevertheless, determining from \sr
experimental measurements the magnitude of the parameters in the
Ginzburg-Landau model is complicated, because the number of
independent parameters
occurring for the case of tetragonal symmetry is greater than for
the case of hexagonal symmetry (i.e. UP$t_3$)
\protect\cite{hes1,lus1,tai1}. Thus for  \sr, three linearly
independent parameters, b$_1$, b$_2$, and b$_3$, are required to
specify the fourth order terms in $\Psi$  occurring in
eqn.~(\ref{G_L_1}) ; whereas only two independent
parameters, b$_1$ and b$_2$, are required for UPt$_3$.
For \sr, two independent ratios can be formed from the
three independent $b_i$ parameters, and these
two independent ratios could be determined, for example, by
experimentally determining the ratios $\Delta C_\sigma^+ / \Delta
C_\sigma^-$ in the presence of the $\sigma_1$ and $\sigma_6$
\protect\cite{con2,con1}.

Measurements results for the \sr elastic constants below T$_c$ are
presented in Ref. \protect\cite{lup2}. There, it is concluded that
the quantities C$_{44}$ and C$_{11}$ - C$_{12}$ follow the same
behavior as those of the BCS superconducting transition, which is
evidenced by a change in slope below T$_{c0}$. On the other hand, a
discontinuity is observed for C$_{66}$ below T$_{c0}$, without a
significant change in the sound speed slope as T goes below 1
Kelvin. It has been previously stated \protect\cite{lup2,con1}
that this kind of C$_{66}$ changes can be understood as a signature of
an unconventional transition to a superconducting phase. Thus, this
set of results and others, as those of Clifford et. al
\protect\cite{clif1}, lead to consider \sr as an excellent candidate
for a detailed experimental investigation of the effects of a
symmetry-breaking field in unconventional superconductors.

\section*{Acknowledgments}

We thank Prof. Michael Walker from the University of Toronto, Prof.
Kirill Samokhin from Brock University, and Prof. Christian Lupien
from Université de Sherbrooke for stimulating discussions. We are
also grateful to the Referee for his comments. This
research was supported by the Grant CDCHTA-ULA number
C-1908-14-05-B.

\end{document}